\documentclass[a4paper,fleqn,usenatbib]{mnras}
\usepackage{graphicx}
\usepackage{newtxtext,newtxmath}
\usepackage{url}
\usepackage[T1]{fontenc}
\newcommand{\dmm}{\ensuremath{\Delta\mu/\mu}}
\newcommand{\kms}{\ensuremath{\textrm{km s}^{-1}}}
\newcommand{\ms}{\ensuremath{\textrm{m s}^{-1}}}
\newcommand{\degs}{\ensuremath{\textrm{deg s}^{-1}}}
\newcommand{\vlsr}{\ensuremath{{\textrm{v}_\textrm{{\scriptsize{LSR}}}}}}
\newcommand{\vtel}{\ensuremath{{\textrm{v}^\textrm{{\scriptsize{tel}}}}}}
\newcommand{\stat}{\ensuremath{\textrm{\scriptsize{stat}}}}
\newcommand{\syst}{\ensuremath{\textrm{\scriptsize{syst}}}}
\newcommand{\obs}{\ensuremath{\textrm{\scriptsize{obs}}}}
\newcommand{\rest}{\ensuremath{\textrm{\scriptsize{rest}}}}
\newcommand{\opt}{\ensuremath{\textrm{\scriptsize{opt}}}}
\newcommand{\rad}{\ensuremath{\textrm{\scriptsize{rad}}}}
\newcommand{\lab}{\ensuremath{\textrm{\scriptsize{lab}}}}
\newcommand{\tkin}{\ensuremath{T_{\textrm{\scriptsize{kin}}}}}
\newcommand{\low}{\ensuremath{\textrm{\scriptsize{low}}}}
\newcommand{\up}{\ensuremath{\textrm{\scriptsize{up}}}}

\def\jms{J. Mol.\ Spectrosc.}

\def\lrr{Living Rev. Relat.}

\def\soviet{Sov. J. Exp. Theor. Phys.}

\def\revmodphys{Rev.\ Mod.\ Phys.}

\title[Testing $\dmm$ with methanol in L1498]{Testing the variability of the proton-to-electron mass ratio from observations of methanol in the dark cloud core L1498}

\author[Dapr\`a et al.]{
M. Dapr\`a$^{1}$, 
C. Henkel$^{2, 3}$, 
S.~A. Levshakov$^{4, 5, 6}$, 
K.~M. Menten$^{2}$, 
S. Muller$^{7}$, 
\newauthor
H.~L. Bethlem$^{1}$,
S. Leurini$^{2, 8}$,
A.~V. Lapinov$^{9}$, 
and W. Ubachs,$^{1}$
\\
$^{1}$Department of Physics and Astronomy, LaserLaB, Vrije Universiteit, De Boelelaan 1081, 1081 HV Amsterdam, The Netherlands\\
$^{2}$Max-Planck-Institut für Radioastronomie, Auf dem Hügel 69, 53121 Bonn, Germany \\
$^{3}$Astronomy Department, King Abdulaziz University, PO Box 80203, 21589 Jeddah, Saudi Arabia \\
$^{4}$A.F. Ioffe Physical-Technical Institute, 194021 St. Petersburg, Russia \\
$^{5}$Electrotechnical University ``LETI'', 197376 St. Petersburg, Russia \\
$^{6}$ITMO University, 191002 St. Petersburg, Russia \\
$^{7}$Department of Space, Earth and Environment, Chalmers University of Technology, Onsala Space Observatory, SE-43992 Onsala, Sweden \\
$^{8}$INAF-Osservatorio Astronomico di Cagliari, Via della Scienza 5, I-09047, Selargius (CA) \\
$^{9}$Institute for Applied Physics, Uljanov Str. 46, 603950 Nizhny Novgorod, Russia
}

\pubyear{2017}

\begin{document}
\label{firstpage}
\pagerange{\pageref{firstpage}--\pageref{lastpage}} 
\maketitle

\begin{abstract}
The dependence of the proton-to-electron mass ratio, $\mu$, on the local matter density was investigated using methanol emission in the dense dark cloud core L1498. Towards two different positions in L1498, five methanol transitions were detected and an extra line was tentatively detected at a lower confidence level in one of the positions. The observed centroid frequencies were then compared with their rest frame frequencies derived from least-squares fitting to a large data set. Systematic effects, as the underlying methanol hyperfine structure and the Doppler tracking of the telescope, were investigated and their effects were included in the total error budget. The comparison between the observations and the rest frame frequencies constrains potential $\mu$ variation at the level of $\dmm < 6 \times 10^{-8}$, at a 3$\sigma$ confidence level. For the dark cloud we determine a total CH$_{3}$OH (A+E) beam averaged column density of \mbox{$\sim 3-4 \times 10^{12}$ cm$^{-2}$} (within roughly a factor of two), an E- to A-type methanol column density ratio of $N$(A-CH$_3$OH)/$N$(E-CH$_3$OH) $\sim 1.00 \pm 0.15$, a density of \mbox{$n$(H$_{2}) = 3 \times 10^{5}$ cm$^{-3}$} (again within a factor of two), and a kinetic temperature of \mbox{$\tkin = 6 \pm 1$ K}. In a kinetic model including the line intensities observed for the methanol lines, the $n$(H$_{2}$) density is higher and the temperature is lower than that derived in previous studies based  on different molecular species; the intensity of the $1_{0} \rightarrow 1_{-1}$ E line strength is not well reproduced.
\end{abstract}


\section{Introduction}
\label{sec:intro_6}
Theories postulating the space-time dependence of fundamental constants, all implying some form of a violation of the Einstein equivalence principle \citep{Uzan2011}, typically invoke additional quantum fields, beyond those of the Standard Model of particle physics, which then couple to the matter or energy density \citep[e. g., ][]{Bekenstein1982,Sandvik2002,Barrow2002}. Such theoretical frameworks may be subdivided into two classes, one connecting to the cosmological scenario of a growing dark energy density, the other connecting to local environmental effects. 

A similar division holds for the observational perspective as well. Variation of fundamental constants, such as the fine structure constant, \mbox{$\alpha = e^2/4 \pi \epsilon_{0} \hbar c$}, and the proton-to-electron mass ratio, \mbox{$\mu = M_{P}/m_{e}$}, on a cosmological time scale is probed by spectroscopy in the early Universe, either via measurements of metal ions \citep{Savedoff1956,Levshakov1994,Dzuba1999,Webb1999,Evans2014}, molecular hydrogen \citep{Thompson1975,Varshalovich1993,Ubachs2016}, ammonia molecules \citep{Flambaum2007,Murphy2008,Henkel2009,Kanekar2011}, or methanol molecules \citep{Bagdonaite2013a,Bagdonaite2013b,Kanekar2015}. On the other hand, the so-called chameleon scenario \citep{Khoury2004,Brax2004} assumes the existence of light scalar fields that acquire effective potential and mass because of their coupling to matter. This phenomenon depends on the local matter density and therefore can be probed through comparisons of different local environments \citep{Brax2014}. As an example, the dependency of the fundamental constants on the gravitational field was investigated via atomic \citep{Berengut2013a} and molecular spectra \citep{Bagdonaite2014b} in the photosphere of white dwarfs. Another example is the coupling of light scalar fields to the local matter density \citep{Olive2008}. Such dependency can be investigated by comparing the measurements of physical properties in the high-density terrestrial environment, \mbox{$n_{\textrm{\scriptsize{Earth}}} \ge 10^{19}$ cm$^{-3}$} and in the comparatively very low-density interstellar clouds, where densities are \mbox{$\ge 14$ orders} of magnitude lower than on Earth.

As in most studies targeting variation of fundamental constants, molecules and their spectra are ideal test grounds. \cite{Levshakov2010a,Levshakov2010b} used the ammonia method, i.e. they compared ammonia, NH$_{3}$, inversion transitions with rotational transitions of molecules that were considered to be co-spatial with ammonia, to investigate the relative difference between the observed proton-to-electron mass ratio,  $\mu_{\obs}$, and the reference laboratory value $\mu_{\lab}$:
\begin{equation}
\frac{\Delta \mu}{\mu} = \frac{\mu_{\obs} - \mu_{\lab}}{\mu_{\lab}}.
\label{eq:dmm_6}
\end{equation}
They derived an upper limit of \mbox{$|\dmm| \le 3 \times 10^{-8}$} at a level of confidence of $3 \sigma$ by observing 41 cold cores in the Galaxy \citep{Levshakov2010a}. More recently, \cite{Levshakov2013} tested this upper limit against potential instrumental errors by employing spectrometers with different spectral resolution. By re-observing 9 cores with the \mbox{Medicina 32m} and the \mbox{Effelsberg 100m} telescopes, they derived a constraint of $|\dmm| < 2 \times 10^{-8}$ at a $3 \sigma$ confidence level.

As already mentioned, the ammonia method relies on the assumption that the emission regions of the considered molecules are co-spatial. This may introduce systematic effects affecting the $\dmm$ value due to chemical segregation. This particularly holds when - as it is commonly being done - NH$_{3}$ line frequencies are coupled to HC$_{3}$N line frequencies. HC$_{3}$N is a molecule representing young, early time chemistry of a molecular cloud, while NH$_{3}$ stands for late time chemistry \citep{Suzuki1992}. The relevant time scales range from several $10^{5}$ to \mbox{$\sim 10^{6}$ yrs}.
       
Methanol, CH$_{3}$OH, is another molecule sensitive to a variation in $\mu$ and it is abundantly present in the Universe, with more than 1000 lines detected in our Galaxy \citep{Lovas2004}. It represents a better target than ammonia for a $\mu$ variation analysis for two main reasons: i) it has several transitions showing higher sensitivities to a varying $\mu$ \citep{Jansen2011,Jansen2011b,Levshakov2011,Jansen2014}, and ii) methanol transitions have different intrinsic sensitivities, therefore it is possible to derive a value for $\dmm$ based only on its transitions, avoiding the assumption of co-spatiality between different molecules as for the ammonia method.

L1498 is a dense core in the Taurus-Auriga complex and represents an example of the simplest environment in which stars form. Its dense gas content was first studied by \cite{Myers1983} and it is considered a starless core, since it is not associated with IRAS \citep{Beichman1986} or \mbox{1.2 mm} point sources \citep{Tafalla2002}. L1498 was identified as a chemically differentiated system by \cite{Kuiper1996} \citep[for extra details about its chemical structure see also ][]{Wolkovitch1997,Willacy1998}. \cite{Tafalla2002} presented line and continuum observations, finding a systematic pattern of chemical differentiation. In a later study, \cite{Tafalla2004} characterised its physical structure and chemical composition, finding that the density distribution of L1498 traced by the \mbox{1.2 mm} dust continuum emission is very close to that of an isothermal sphere, with a central density of \mbox{$n \sim 10^{5}$ cm$^{-3}$}. They also reported that the gas temperature distribution in the core is consistent with a constant value of \mbox{$\tkin = 10 \pm 1$ K} and that lines have a constant, non-thermal full width at half maximum, \mbox{\emph{FWHM}, of $\Delta V \sim 0.125$ \kms} in the inner core. More recently, \cite{Tafalla2006} investigated the molecular emission of 13 species, including methanol, in L1498. They found that the abundance profiles of most species suffer from a significant drop towards the core centre \citep[see \mbox{Fig. 1} of ][]{Tafalla2006}, resulting in a ring-like distribution around the central dust peak indicating depletion of these molecules onto icy dust grain mantles (`freeze out'). They showed that methanol emission forms a ring with two discrete peaks, one to the SE (hereafter L1498-1) and one to the NW (hereafter L1498-2) of the dust peak. It is noted that the NW methanol peak is dimmer, with a peak intensity of $\sim 80\%$ of the SE peak intensity.

In view of its low $\tkin$ and low degree of turbulence \citep[non-thermal \emph{FWHM} $\simeq 0.12$ \kms, ][]{Tafalla2004}, L1498 shows narrow emission lines and therefore it is an ideal target for measuring a $\mu$-variation. This dense core was already included in the sample studied by \cite{Levshakov2010a,Levshakov2010b,Levshakov2013}. In the following, the detection of 5 methanol lines in L1498-1 and L1498-2 using the Institut de Radio Astronomie Millim\'etrique, IRAM, 30m telescope\footnote{Based on observations carried out under project number 014-14 with the IRAM 30m Telescope. IRAM is supported by INSU/CNRS (France), MPG (Germany) and IGN (Spain).}is presented and it is used to derive a value for $\dmm$ to test the chameleon scenario and to constrain physical parameters of the cloud. The observations are described in \mbox{Section \ref{sec:obs_6}}, and their results in \mbox{Section \ref{sec:results_6}}. The physical parameters of L1498 are derived in \mbox{Section \ref{sec:cloud_6}}, while the laboratory rest frequencies used in this work are discussed in \mbox{Section \ref{sec:restframe_6}}. Finally, the measurement of $\dmm$ is presented in \mbox{Section \ref{sec:dmm_6}}, and the results obtained are summarised in \mbox{Section \ref{sec:conclusion_6}}.

\section{Observations}
\label{sec:obs_6}
The IRAM 30m observations of L1498 were carried out on July 20th and 21st 2014 using the Eight MIxer Receiver \citep[EMIR,][]{Carter2012}. Two EMIR setups were used throughout the observations to target the methanol transitions. The \mbox{3 mm} E090 band was used to measure methanol lines in the range 96.7-\mbox{109 GHz}, while lines in the range 145-\mbox{157.3 GHz} were observed using the \mbox{2 mm} E150 band. For both setups, the EMIR lower and upper inner sidebands, LI and UI respectively, were used in dual-polarisation mode. The data were recorded using the VErsatile SPectrometer Array (VESPA) with its \mbox{3 kHz} channel width, corresponding to velocity channels of $\sim 10$ and \mbox{7 \ms} for the E090 and E150 bands, respectively. 

The observations were performed using the position switching mode with offsets of $\pm 600$, $\pm 900$, or \mbox{$\pm 1800$ arcsec} in azimuth. Alternating between western and eastern offsets helped to provide flat spectral baselines. Comparing data with different offset positions did not reveal any significant differences, thus ensuring that emission from the offset positions did not affect the resulting spectra. A single scan's duration was $\sim 6$ min, divided into two parts. The first half involved integration in the off-source position and the second half was devoted to the on-source integration. Sky frequencies were updated at the beginning of each scan (see \mbox{Section \ref{subsubsec:doppler_6}}). Pointing was checked frequently and was found to be stable within 3 arcsec. Calibration was obtained every 12 minutes using standard hot/cold-load absorber measurements. The data reduction was performed using the \textsc{gildas} software's \textsc{class} package\footnote{\url{http://www.iram.fr/IRAMFR/GILDAS/}}. Only linear baselines were subtracted from the individual spectra. The raw frequencies were Doppler shifted in order to correct for Earth motion relative to L1498 during the observations \citep[$\vlsr = 7.8$ \kms,][]{Tafalla2004}. The same value of $\vlsr$ was used for the two observed positions.

The IRAM 30m telescope adopts the optical definition of radial velocity, while the radio definition is implemented in \textsc{class}. The optical radial velocity is defined as:
\begin{equation}
    \frac{\textrm{v}^{\opt}_{\textrm{\scriptsize{LSR}}}}{c} = \frac{\nu_{\rest}}{\nu_{\obs}} - 1,
    \label{eq:optvel_6}
\end{equation}
where $\nu_{\obs}$ is the observed frequency and $\nu_{\rest}$ is the reference frequency in the rest frame. In contrast, the radio definition of the radial velocity is given by:
\begin{equation}
    \frac{\textrm{v}^{\rad}_{\textrm{\scriptsize{LSR}}}}{c} = 1 - \frac{\nu_{\obs}}{\nu_{\rest}},
    \label{eq:radvel_6}
\end{equation}
using the same symbols as in \mbox{Eq. (\ref{eq:optvel_6})}. As a consequence, the radial velocities returned by the two definitions are different, $\textrm{v}_{\opt} \ne \textrm{v}_{\rad}$. For L1498, which has \mbox{$\vlsr = 7.8$ \kms}, the discrepancy between the two definitions is of the order of \mbox{$\sim 10^{-8}$ \kms} and hence is considered negligible.

The observation time was divided unevenly between the two peaks, resulting in \mbox{$\sim 14$ hrs} of integration on L1498-1 and \mbox{$\sim 4$ hrs} on L1498-2. This was done in order to reach a good signal-to-noise ratio, S/N, for the source with the brighter emission \citep[the SE peak L1498-1, see][]{Tafalla2006}.\\

\section{Results}
\label{sec:results_6}
\begin{figure*}
    \centering
    \includegraphics[width=2\columnwidth]{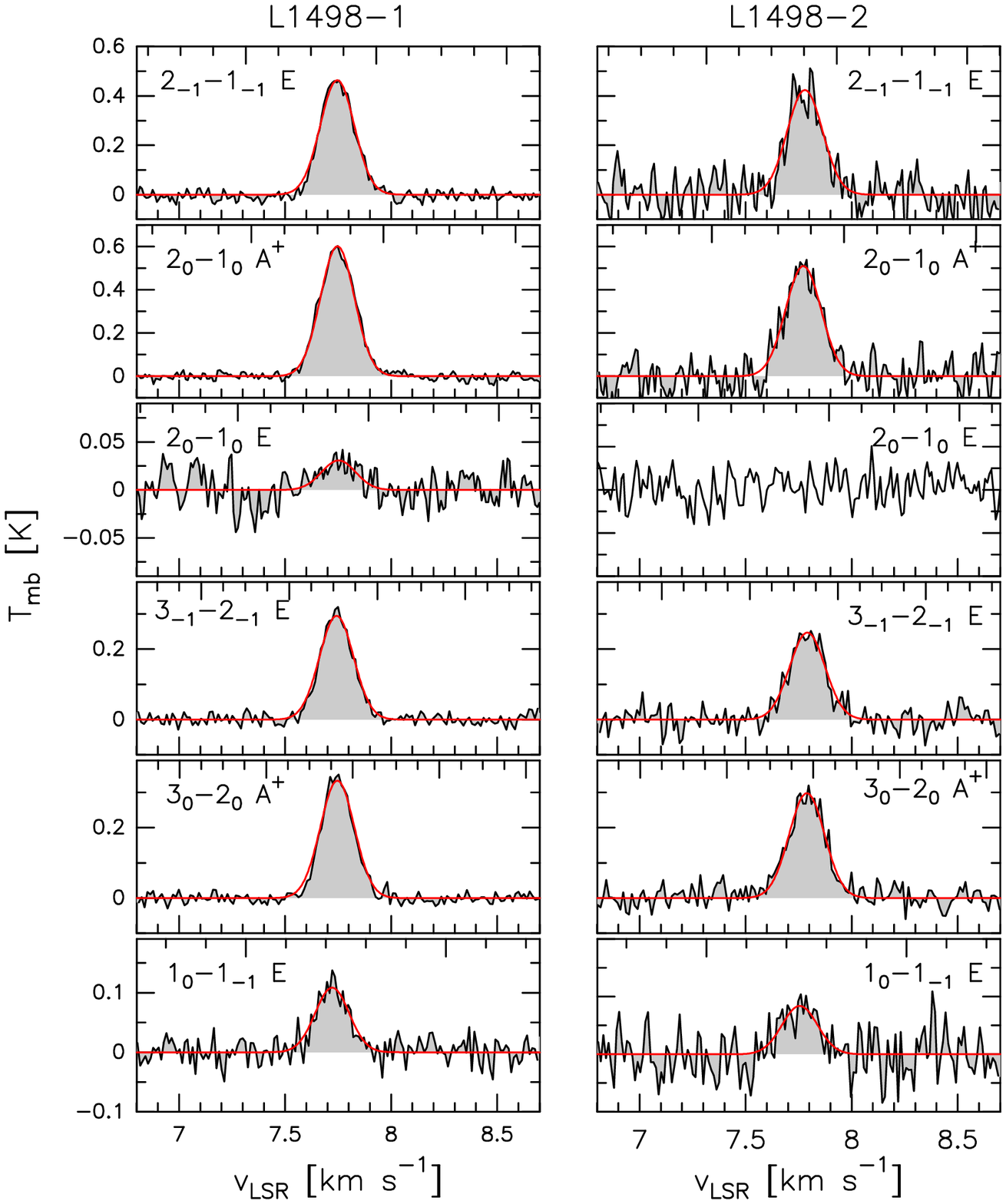}
    \caption{Spectra of the detected methanol transitions in L1498-1 (left panels; $\alpha_{\scriptsize{\textrm{J}2000}}$ = 04$^{\rm h}$ 10$^{\rm m}$ 56.6$^{\rm s}$, $\delta_{\scriptsize{\textrm{J}2000}}$ = +25$^{\circ}$ 09$'$ 08$''$) and L1498-2 (right panels; $\alpha_{\scriptsize{\textrm{J}2000}}$ = 04$^{\rm h}$ 10$^{\rm m}$ 47.0$^{\rm s}$, $\delta_{\scriptsize{\textrm{J}2000}}$ = +25$^{\circ}$ 10$'$ 18$''$). The (red) solid lines show the Gaussian fits. The transition \mbox{$3_{-1} \rightarrow 2_{-1}$ E} fell near the edge of the sideband, hence the background is not provided beyond radial velocities of \mbox{$\sim 8.8$ \kms}. The \mbox{$2_{0} - 1_{0}$ E} transition was not detected towards the L1498-2 position.}
    \label{fig:fits_6}
\end{figure*}

Several methanol transitions were detected at a $> 10 \sigma$ confidence level in L1498-1 and L1498-2, respectively. All the detected lines are well described by a single Gaussian profile, as shown in \mbox{Fig. \ref{fig:fits_6}}, which has three fitting parameters: the line intensity, expressed in terms of the main beam brightness temperature $T_{\textrm{\scriptsize{mb}}}$, the line width \emph{FWHM}, and the transition frequency $\nu_{\obs}$, which is corrected for the shift introduced by the radial velocity of L1498 \citep[\mbox{$\vlsr \sim 7.8$ \kms},][]{Tafalla2006}. The line parameters related to each detected transition in L1498-1 and L1498-2 are listed in \mbox{Table \ref{tab:parameters_6}}. All the detected methanol transitions turn out to be optically thin.

The weak transition \mbox{$2_{0} \rightarrow 1_{0}$ E} (at a rest-frame frequency of \mbox{$\nu_{\rest} \simeq 96.744545$ GHz}) was detected at a confidence level of $5 \sigma$ in the brighter position L1498-1 only. Although weak, it is well described by a single Gaussian profile, as shown in \mbox{Fig. \ref{fig:fits_6}}, whose parameters are included in \mbox{Table \ref{tab:parameters_6}}. An extra line, the \mbox{$2_{0} \rightarrow 2_{-1}$ E} transition (\mbox{$\nu_{\rest} \simeq157.276023$ GHz}), may have also been detected with a confidence level of $3 \sigma$ in L1498-1. However, the detection is tentative. The line may be broadened by the noise and is hence not considered further in this work.

\section{Cloud properties}
\label{sec:cloud_6}
As already briefly mentioned in \mbox{Section \ref{sec:intro_6}}, \cite{Tafalla2004} determined physical and chemical characteristics of the dark cloud L1498, based on measurements of NH$_3$, N$_2$H$^{+}$, CS, C$^{34}$S, C$^{18}$O and C$^{17}$O. Our CH$_{3}$OH line data can be used to complement their results. To simulate the measured line parameters we applied the \textsc{radex} non-LTE code model \citep{Schoier2005,vanderTak2007}, which is based on collision rates with H$_{2}$ reported by \cite{Rabli2010}, using kinetic temperature, density, and E- or A-type methanol column density as variables. The code calculates line intensities for a uniform sphere, which is justified because of the overall shape of the cloud and the radial profile derived from ammonia for the kinetic temperature \citep{Tafalla2004}, which resulted in a constant value. For the background temperature we adopted 2.73 K and for the line width, being relevant for the methanol column density, an average value derived from \mbox{Table \ref{tab:parameters_6}} (\mbox{$\Delta V = 0.165$ \kms}; note that this is slightly larger than the value taken from the literature and given in \mbox{Section \ref{sec:intro_6}}). Since the line width is known to a far higher accuracy than the column density (the latter is only known to a factor of roughly two), the specific choice of $\Delta V$ is not critical. With this model most line intensities are well reproduced. This also includes our non-detection of the \mbox{$2_0 \rightarrow 2_{-1}$ E} transition, with the model mostly indicating absorption at line intensities below our sensitivity limit. Nevertheless, none of the fits is perfect, because we do not obtain good simulations of the \mbox{$1_0 \rightarrow 1_{-1}$ E} transition. For this transition the modelled intensities are too low by factors of $\sim 2-3$.

\begin{figure}
    \centering
    \includegraphics[width=\columnwidth, angle=-90]{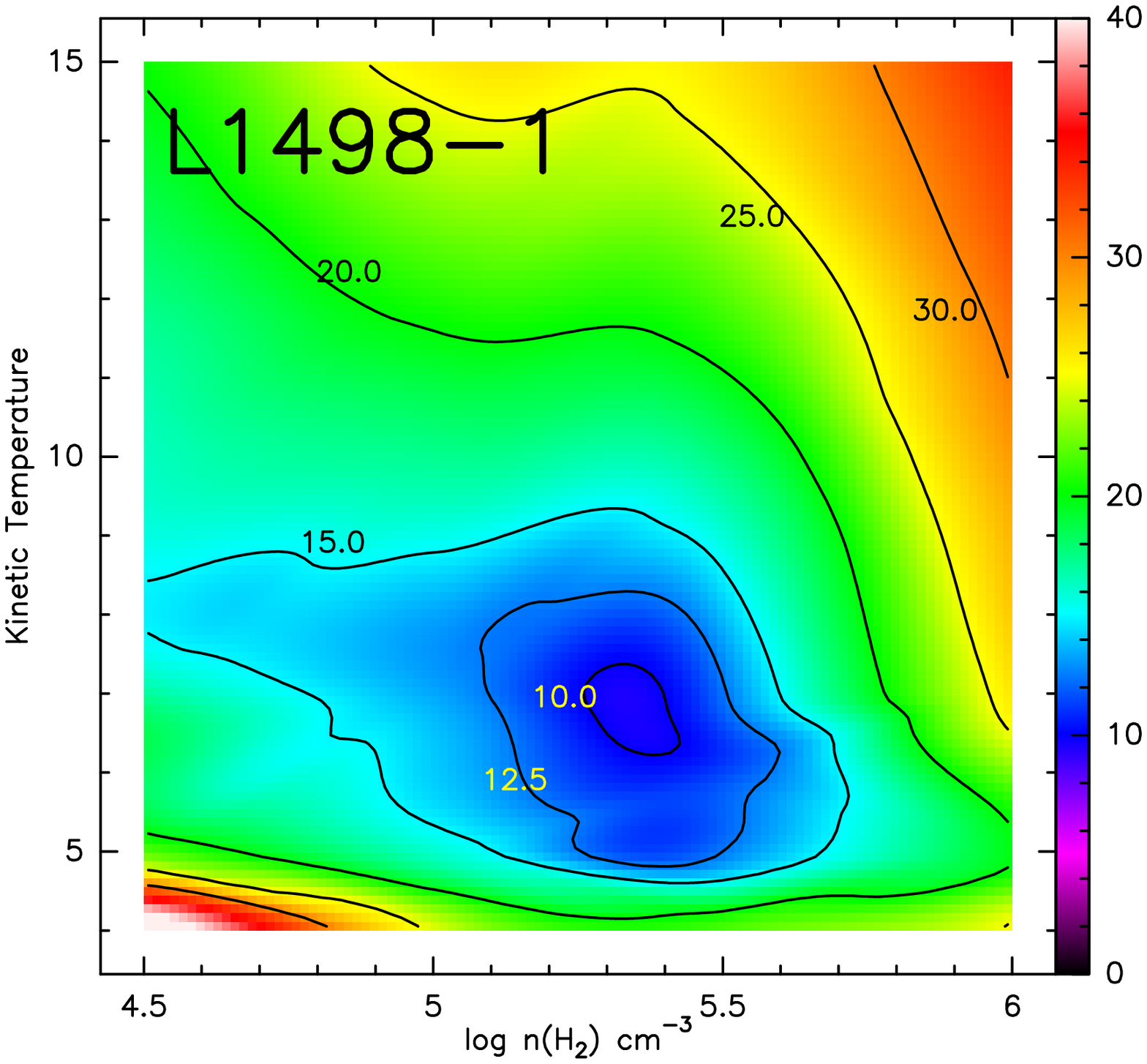}
    \includegraphics[width=\columnwidth, angle=-90]{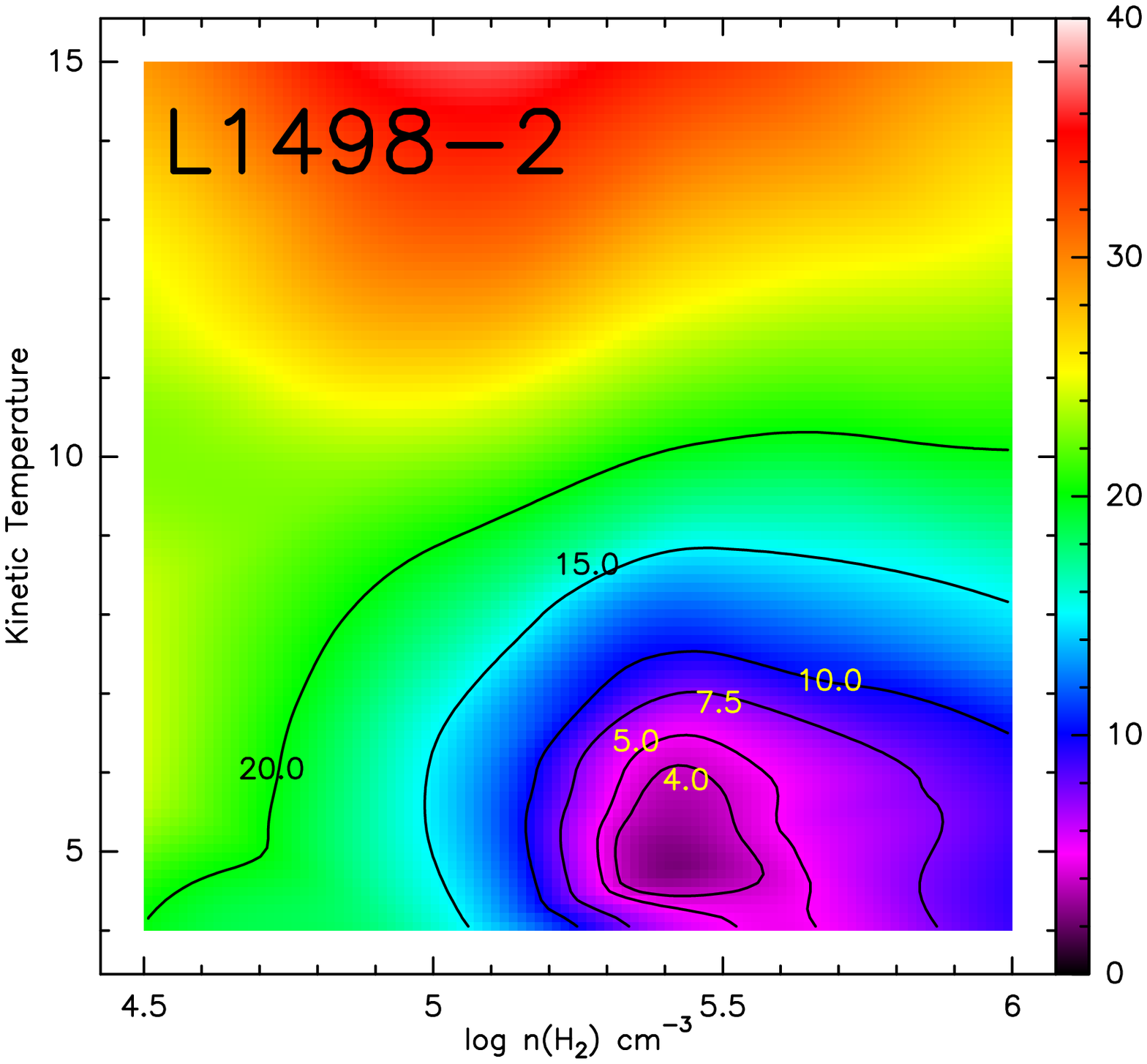}
    \caption{Map of the reduced $\chi^{2}$ value as function of density and kinetic temperature values used to simulate the E- and A-type main beam brightness temperatures in L1498-1 (upper panel) and L1498-2 (lower panel).}
    \label{fig:chi2_1}
\end{figure}

To find the best solution(s), we used \textsc{radex} to create a grid in $\tkin$ and $n(\textrm{H}_{2})$, optimising for each ($T_{\rm kin}$, $n(\textrm{H}_{2})$) pair the column density by calculating reduced $\chi^2$ values. With six (L1498-1) and five (L1498-2) transitions and three free parameters (kinetic temperature, density, and column density), there are three and two degrees of freedom, respectively. Because of the above mentioned \mbox{$1_0 \rightarrow 1_{-1}$ E} line, reduced $\chi^{2}$ values, i.e the sum of the squared differences between observed and modelled line intensities, divided by the square of the uncertainty in the measured parameter (10\% of the line temperature and standard deviation of the Gaussian amplitude, added in quadrature) never get close to unity. Nevertheless, \mbox{Fig. \ref{fig:chi2_1}} provides resulting values for L1498-1 and L1498-2 in the two dimensional ($\tkin$, $n(\textrm{H}_{2})$) space with background shading and contours reflecting the reduced $\chi^2$ values. For L1498-1, the source with best determined line intensities, the best solution yields a column density of order \mbox{$N$(E+A methanol) $\sim 3 \times 10 ^{12}$ cm$^{-2}$}, \mbox{$\tkin \sim 6$ K}, and \mbox{$n(\textrm{H}_{2}) \sim 3 \times 10^5$ cm$^{-3}$}. For position L1498-2, the optimal solution returns a similar density, a lower value for $\tkin$ of \mbox{$\sim 5$ K} and a methanol column density of \mbox{$N$(E+A methanol) $\sim 4 \times 10 ^{12}$ cm$^{-2}$}. In other words, the gas traced by methanol appears to be cooler and denser than that analysed by \cite{Tafalla2004}, but the temperatures are close to the value of \mbox{$7.1 \pm 0.5$ K} reported by \cite{Levshakov2010b}. The results returned by the \textsc{radex} code are presented in \mbox{Column 6} of \mbox{Table \ref{tab:parameters_6}}.
\begin{table*}
    \centering
    \caption{Details of the methanol transitions observed in L1498. Columns 3-4 show the measured frequencies and line widths. Columns 5-6 present the measured and modeled (main beam) brightness temperatures, the latter for the solution with the lowest reduced $\chi^2$ value. The optical depths $\tau_{\nu}$, calculated with the \textsc{lvg} code, are listed in column 7. Columns 8-9 present the energies of the upper and lower states above the ground state. Column 10 shows the sensitivities of the transitions to a varying proton-to-electron mass ratio and column 11 presents the EMIR band in which the transitions were observed. The \textsc{lvg} simulations were performed assuming a molecular hydrogen density of \mbox{$n(\textrm{H}_{2}) = 3 \times 10^{5}$ cm$^{-3}$}, a kinetic temperature of \mbox{$\tkin = 6$ K}, and an E- and A-type methanol column density of \mbox{$N_{\textrm{\scriptsize{E}}} = N_{\textrm{\scriptsize{A}}} = 3.3\times 10^{12}$ cm$^{-2}$} for L1498-1. For simulating the methanol transitions in L1498-2, a molecular hydrogen density of \mbox{$n(\textrm{H}_{2}) = 3 \times 10^{5}$ cm$^{-3}$}, a kinetic temperature of \mbox{$\tkin = 5$ K}, and an E- and A-type methanol column density of \mbox{$N_{\textrm{\scriptsize{E}}} = N_{\textrm{\scriptsize{A}}} = 4.3\times 10^{12}$ cm$^{-2}$} were assumed.}
    \label{tab:parameters_6}
    \begin{tabular}{clrcrccrrcc}
    \hline
     Source & Transition	& \multicolumn{1}{c}{Observed frequencies$^{a}$}	& FWHM	& \multicolumn{1}{c}{T$_{\textrm{mb}}$}		& T$_{\textrm{\textsc{radex}}}$	& $\tau_{\nu}$	& \multicolumn{1}{c}{$E_{\up}$}		& \multicolumn{1}{c}{$E_{\low}$}	& Sensitivity	& Band	\\
     		&			& \multicolumn{1}{c}{[MHz]}					& [MHz]	& \multicolumn{1}{c}{[mK]}					& [mK]					& 				& \multicolumn{1}{c}{[K]}			& \multicolumn{1}{c}{[K]}			& coefficients	& 		\\
    \hline
    \hline
    L1498-1				& $2_{-1} \rightarrow 1_{-1}$ E		& $96739.376 \pm 0.001$		& $0.046 \pm 0.003$	& $556 \pm 15$& 524			& 0.23		& 12.2		& 7.7			& -1.0		& E090	\\
    $\alpha = ~~04:10:56.6$	& $2_{0} \rightarrow 1_{0}$ A$^{+}$	& $96741.388 \pm 0.001$		& $0.049 \pm 0.003$	& $716 \pm 15$& 620			& 0.27		& 6.8			& 2.3		& -1.0	& 			\\
    $\delta = +25:09:08.0$		& $2_{0} \rightarrow 1_{0}$ E		& $96744.560 \pm 0.007$		& $0.055 \pm 0.003$	& $32 \pm 17$	& 60				& 0.03		& 19.5		& 15.0		& -1.0		& 		\\
    						& $3_{-1} \rightarrow 2_{-1}$ E		& $145097.464 \pm 0.001$	& $0.085 \pm 0.001$	& $374 \pm 14$& 412			& 0.23		& 19.0		& 12.2		& -1.0		& E150	\\
    						& $3_{0} \rightarrow 2_{0}$ A$^{+}$	& $145103.212 \pm 0.001$	& $0.077 \pm 0.001$	& $426 \pm 12$& 423			& 0.23		& 13.6		& 6.8			& -1.0		& 		\\
    						& $1_{0} \rightarrow 1_{-1}$ E		& $157270.877 \pm 0.003$	& $0.079 \pm 0.005$	& $137 \pm 19$&  94			& 0.44		& 15.0		& 7.7			& -3.5		& 		\\
    						& $2_{0} \rightarrow 2_{-1}$ E		& \multicolumn{3}{c}{Not detected}							&  52				& 0.32		& 19.5		& 12.2		& -3.5		& 		\\
    \hline
    L1498-2				& $2_{-1} \rightarrow 1_{-1}$ E		& $96739.365 \pm 0.002$			& $0.048 \pm 0.003$	& $508 \pm 75$	& 545	& 0.37		& 12.2		& 7.7		& -1.0		& E090	\\
    $\alpha = ~~04:10:47.0$	& $2_{0} \rightarrow 1_{0}$ A$^{+}$	& $96741.380 \pm 0.001$			& $0.054 \pm 0.003$	& $608 \pm 69$	& 619	& 0.42		& 6.8			& 2.3		& -1.0		& 		\\
    $\delta = +25:10:18.0$		& $2_{0} \rightarrow 1_{0}$ E		& \multicolumn{3}{c}{Not detected}									& 60		& 0.04		& 19.5		& 15.0		& -1.0	& 		\\
    						& $3_{-1} \rightarrow 2_{-1}$ E		& $145097.440 \pm 0.001$		& $0.097 \pm 0.005$	& $325 \pm 29$	& 360	& 0.36		& 19.0	& 12.2		& -1.0		& E150	\\
    						& $3_{0} \rightarrow 2_{0}$ A$^{+}$	& $145103.191 \pm 0.002$		& $0.091 \pm 0.003$	& $372 \pm 26$	& 357	& 0.30		& 13.6	& 6.8			& -1.0		& 		\\
    				 		& $1_{0} \rightarrow 1_{-1}$ E		& $157270.858 \pm 0.010$		& $0.100 \pm 0.013$	& $115 \pm 38$		&  88		& 0.66		& 15.0	& 7.7			& -3.5		& 		\\
    						& $2_{0} \rightarrow 2_{-1}$ E		& \multicolumn{3}{c}{Not detected}									&  41		& 0.41		& 19.5	& 12.2	& -3.5	& 		\\
    \hline
    \multicolumn{11}{l}{$^{a}$ corrected for the radial velocity of L1498}
    \end{tabular}
\end{table*}

This kinetic temperature is lower than the lower limit reported by \cite{Goldsmith1978}, who claimed that there is no equilibrium solution for \mbox{$\tkin < 8$ K} in their model of thermal balance in dark clouds. Assuming that cosmic rays are the only heating source of the dark cloud, their model returns gas temperatures in the range \mbox{8-12 K}. A similar result was obtained by \cite{Goldsmith2001}, who used an \textsc{lvg} model for radiative transfer including the effect of the gas-dust coupling at different depletions of the major molecular coolants. 

While a reduced $\chi^2$ value of order 10, mainly due to the \mbox{$1_0 \rightarrow 1_{-1}$ E} line, in L1498-1 and of order 3 in L1498-2 may raise doubts about the above mentioned results, it is nevertheless a result worth to be presented and to be confirmed (or rejected) by other methanol transitions as well as by lines from other molecular species. A possible explanation for the high values of the reduced $\chi^{2}$ may lie in the uncertainties in the collisional rates used in \textsc{radex} and in our limited knowledge on cloud geometry and 3-dimensional velocity field. However, testing this is beyond the goal of this work. Another remarkable result is that \emph{N}(E-CH$_3$OH$) \sim N$(A-CH$_3$OH), with an accuracy of \mbox{$\sim15\%$}. Here we should consider that the lowest E-state, the \mbox{$1_{-1}$ E} level, is \mbox{7.9 K} above the lowest $(0_0)$ A-state. Following \cite{Friberg1988}, the E/A abundance ratio for a thermalisation temperature of \mbox{10 K} becomes 0.69, which is as far from unity as our calculations may (barely) permit. For a thermalisation at \mbox{6 K}, our determined E/A abundance ratios would then clearly be too large. Perhaps, the bulk of the methanol has been formed prior to the occurrence of highly obscured deeply shielded regions with kinetic temperatures below \mbox{10 K}.

\section{Rest frequencies}
\label{sec:restframe_6}
The methanol lines detected in L1498 are very narrow, with a width of less than \mbox{$\sim 0.2$ \kms} corresponding to \mbox{$\sim 50$ kHz} for the lines observed at \mbox{96 GHz} and \mbox{$\sim 80$ kHz} for the lines observed near \mbox{150 GHz} (see \mbox{Table \ref{tab:parameters_6}}). In order to derive a tight constraint on a varying proton-to-electron mass ratio, the astrophysical data need to be compared with the most accurate values for the rest frequencies. In the following, a concise review on the laboratory frequencies is provided.

The main body of laboratory frequencies for methanol lines stems from the study of \cite{Lees1968} and the subsequently published database \citep{Lees1973}. The lines presently observed in L1498 are all contained in that list and are reproduced in \mbox{Table \ref{tab:rest_freq_6}}. The estimated uncertainty for these laboratory measurements is about \mbox{50 kHz} for all transitions. Later, transition frequencies at a much higher accuracy were measured for a limited number of lines by beam-maser spectroscopy \citep{Radford1972,Heuvel1973,Gaines1974} and by microwave Fourier-transform spectroscopy \citep{Mehrotra1985}. None of these measurements refer to the transitions considered in this work.
\begin{table*}
    \centering
    \caption{List of rest-frame frequencies derived from laboratory data and numerical calculations.}
    \label{tab:rest_freq_6}
    \begin{tabular}{lrrr}
    \hline
    Transition	& \multicolumn{2}{c}{Measured frequency}	& \multicolumn{1}{c}{Calculated}		\\%& Sensitivity \\
     		  	& \multicolumn{2}{c}{[MHz]}				& \multicolumn{1}{c}{frequency [MHz]}	\\%& coefficients \\
    \hline
    \hline
    $2_{-1} \rightarrow 1_{-1}$ E		&   96739.390(50)$^{a}$	& 96739.362(5)$^{b}$	& 96739.358(2)$^{d}$	\\%& $-1.0$	\\
    $2_{0} \rightarrow 1_{0}$ A$^{+}$	&   96741.420(50)$^{a}$	& 96741.375(5)$^{b}$	& 96741.371(2)$^{d}$	\\%& $-1.0$	\\
    $2_{0} \rightarrow 1_{0}$ E		&   96744.580(50)$^{a}$	& 96744.550(5)$^{b}$	& 96744.545(2)$^{d}$	\\%& $-1.0$	\\
    $3_{-1} \rightarrow 2_{-1}$ E		& 145097.470(50)$^{a}$	& 145097.370(10)$^{c}$	& 145097.435(3)$^{d}$	\\%& $-1.0$	\\
    $3_{0} \rightarrow 2_{0}$ A$^{+}$	& 145103.230(50)$^{a}$	& 145103.152(10)$^{c}$	& 145103.185(3)$^{d}$	\\%& $-1.0$	\\
    $1_{0} \rightarrow 1_{-1}$ E		& 157270.700(50)$^{a}$	& 157270.851(10)$^{c}$ 	& 157270.832(12)$^{d}$	\\%& $-3.5$	\\
    \hline
    \multicolumn{4}{l}{$^{a}$from \cite{Lees1968}} \\
    \multicolumn{4}{l}{$^{b}$from \cite{Muller2004}} \\
    \multicolumn{4}{l}{$^{c}$from \cite{Tsunekawa1995}} \\
    \multicolumn{4}{l}{$^{d}$from \cite{Xu2008}} \\
    \end{tabular}
\end{table*}

\cite{Hougen1994} developed the \textsc{belgi} code, which is a program to calculate and fit transitions of molecules containing an internally hindered rotation. This program was employed by \cite{Xu1997} to produce a fit of 6655 molecular lines delivering the values of 64 molecular constants describing the rotational structure of methanol. Based on these molecular parameters the transition frequencies can be recalculated with a higher accuracy, assuming that the fitting procedure averages out the statistical uncertainties in the experimental data. 

In 2004, \cite{Muller2004} published a study focusing on the laboratory measurements of methanol lines that can be observed either in masers or in dark clouds. In their studies, they used the Cologne spectrometer, in the range \mbox{60-119 GHz}, and the Kiel spectrometer, in the range \mbox{8-26 GHz}. The three lines at \mbox{96 GHz}, presently observed in L1498, were measured with an estimated accuracy of \mbox{5 kHz}. \cite{Muller2004} also listed transition frequencies for four lines at 145 and \mbox{157 GHz}, that were previously measured with a claimed accuracy of \mbox{10 kHz} by \cite{Tsunekawa1995}. These most accurate experimental values for the presently observed lines are also listed in \mbox{Table \ref{tab:rest_freq_6}}.

More recently, \cite{Xu2008} improved the theoretical analysis based on an extended version of the \textsc{belgi} program, with the inclusion of  a large number of torsion-rotation interaction terms to the Hamiltonian, and by fitting a huge data set including $\sim 5600$ frequency measurements in the microwave, sub-millimetre wave and Terahertz regime, as well as a set of $19~000$ Fourier transform far infrared wavenumber measurements to 119 molecular parameters. By replacing some older measurements, which exhibit larger residuals, with more accurate frequency measurements obtained since 1968, they derived an improved set of rest frequencies\footnote{\url{https://spec.jpl.nasa.gov/ftp/pub/catalog/catdir.html}} that is included in the last column of \mbox{Table \ref{tab:rest_freq_6}}.

Inspection of \mbox{Table \ref{tab:rest_freq_6}} shows that a unique and consistent data set of rest frame frequencies does not exists for the six lines of relevance to our observations. Moreover it is clear that the observations in the cold L1498 core are more accurate than most experimental data from the laboratory, and for most lines are also more accurate than the results from the least-squares fitting treatment based on the \textsc{belgi} model. For the three lines at \mbox{96 GHz} the laboratory measurements performed with the K\"{o}ln spectrometer \citep{Muller2004}, with an estimated uncertainty of \mbox{5 kHz}, agree very well with the data resulting from the extended \textsc{belgi} analysis that have an estimated uncertainty of \mbox{2 kHz} \citep{Xu2008}. The deviations between the data sets are \mbox{$\leq 5$ kHz}. The frequencies of these lines are some \mbox{40-50 kHz} lower than in the old measurements of \cite{Lees1968}, but in that study the accuracy was estimated at \mbox{50 kHz}, so that the older data agree with the more recent and more accurate data within $1\sigma$. From this we conclude that for the 96 GHz lines we have a consistent set of rest frame frequencies to an accuracy well below our already rather narrow channel spacing of \mbox{10 kHz}. Therefore, the most accurate set represented by the values from the extended \textsc{belgi} analysis of \cite{Xu2008} was chosen as reference.

For the two higher frequency transitions at \mbox{145 GHz} the situation is less clear. Experimental laboratory data stem from the Toyama atlas \citep{Tsunekawa1995} for which a measurement uncertainty of 10 kHz was claimed. These data deviate with 65 kHz and 33  kHz from the extended \textsc{belgi} analysis. In a 2008 updated least-squares analysis an uncertainty of \mbox{3 kHz} is specified. In comparison to the  older measurements of \cite{Lees1968} the calculated values  \citep{Xu2008} are at the limit of $1 \sigma$ deviation while the deviation from the Toyama data amounts to $2 \sigma$. 

For the line at \mbox{157 GHz} the Toyama value agrees well within combined error margins with the calculated value. However both values for this line deviate by $3 \sigma$ from the older value measured by \cite{Lees1968}. In view of the fact that for the other five lines the theoretical value is expected to be most accurate, for consistency we will also include the extended \textsc{belgi} analysis of \cite{Xu2008} for the \mbox{157 GHz} line in the rest frame set. 

\begin{figure}
    \centering
    \includegraphics[width=\columnwidth]{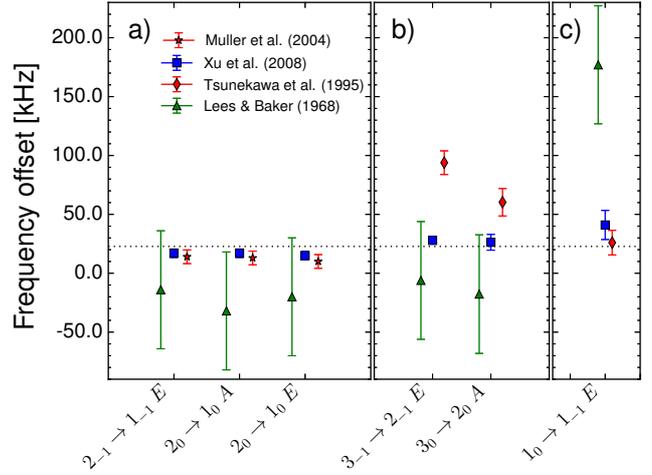}
    \caption{Frequency offsets between the observed, velocity corrected (see \mbox{Section \ref{sec:restframe_6}}) frequencies and the rest frame frequencies for the methanol transitions detected in L1498-1 at $\sim 96$ GHz (Panel a), $\sim 145$ GHz (Panel b) and $\sim 157$ GHz (Panel c). The (blue) squares show the comparison with the frequencies reported by \protect \cite{Xu2008}, the (green) triangles  the frequencies measured by \protect \cite{Lees1968}, the (red) stars and the (red) diamonds the frequencies reported by \protect \cite{Muller2004} and \protect \cite{Tsunekawa1995}, respectively. The dotted line at $\sim 22$ kHz shows the average deviation of the preferred rest frequency data from the observations.}
    \label{fig:consistency_6}
\end{figure}

From the above evaluation and the graphical comparison of data sets presented in \mbox{Fig. \ref{fig:consistency_6}} we decide that the results from the comprehensive least-squares fitting treatment \citep{Xu2008} provides the best and most consistent set of rest frequencies, in agreement with the accurate data from \cite{Muller2004}, the 157 GHz line measured by \cite{Tsunekawa1995}, and with the older measurements reported by \cite{Lees1968}. Future improved laboratory measurements may decide on the validity of this choice.

\begin{figure}
    \centering
    \includegraphics[width=\columnwidth]{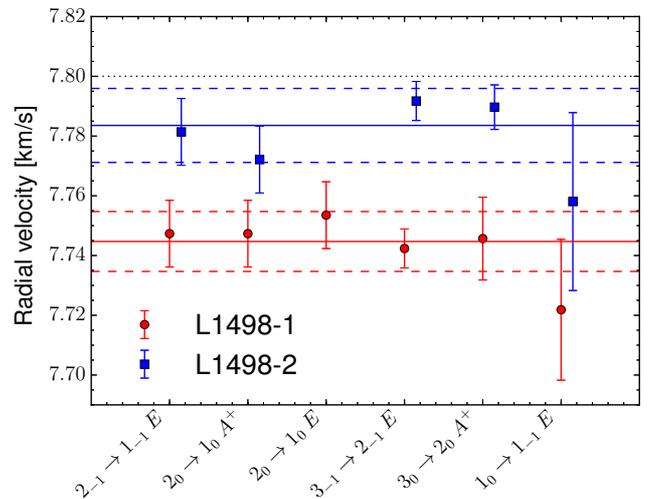}
    \caption{Radial LSR velocities for positions L1498-1 (red circles) and L1498-2 (blue squares) derived by the comparison of the observed frequencies with those calculated using the extended version of the \textsc{belgi} program \protect \citep{Xu2008}. The solid line represent the radial velocity of the two positions, while the dashed lines show their $\pm 1 \sigma$ boundaries. The (black) dotted line shows the $\vlsr$ reported by \protect \cite{Tafalla2004}.}
    \label{fig:vel_offsets_6}
\end{figure}

The weighted average of the deviations between the observed and the \textsc{belgi} frequencies returns values of $\Delta \nu_{1} = \nu_{\obs} - \nu_{\lab} = 22 \pm 9$ and \mbox{$\Delta \nu_{2} = 7 \pm 6$ kHz} for L1498-1 and L1498-2, respectively. These deviations were translated into radial velocities using the radio definition, as in \mbox{Eq. (\ref{eq:radvel_6})}, delivering $\vlsr_{1} =  7.745 \pm 0.010$ and \mbox{$\vlsr_{2} = 7.784 \pm 0.012$ \kms} for position L1498-1 and L1498-2 respectively, as shown in \mbox{Fig. \ref{fig:vel_offsets_6}}. We deduce from this that the two positions L1498-1 and L1498-2 have different radial velocities.
%The two velocities do not agree with the value of \mbox{$\vlsr = 7.8$ \kms} reported by \cite{Tafalla2004} and they do not match within their errors. 

As will be discussed in the next section, the analysis of the presently observed highly accurate astronomical data, exhibiting an accuracy of \mbox{5 kHz} or better (see \mbox{Table \ref{tab:parameters_6}}), will provide a consistency check on the choice of the preferable, most accurate rest frame data set.

\section{Constraining $\dmm$}
\label{sec:dmm_6}
The variation of the proton-to-electron mass ratio was investigated using the methanol transitions that were detected in the two emission peaks of the dark cloud L1498. The observed frequencies of the selected transitions were compared with the rest frame frequencies reported by \cite{Xu2008}, as discussed in \mbox{Section \ref{sec:restframe_6}}, and were subsequently interrelated via:
\begin{equation}
    \frac{\Delta \nu_{i}}{\nu} = K_{i} \frac{\Delta \mu}{\mu},
    \label{eq:shift_6}
\end{equation}
where $\Delta \nu_{i}/\nu = (\nu_{\obs} - \nu_{\lab})/\nu_{\lab}$ is the relative difference between the observed and the calculated frequency, of the \emph{i}-th transition, $K_{i}$ its sensitivity coefficient, and $\dmm$ is the relative difference between the proton-to-electron mass ratio in the dark cloud and on Earth.

\begin{figure}
    \centering
    \includegraphics[width=\columnwidth]{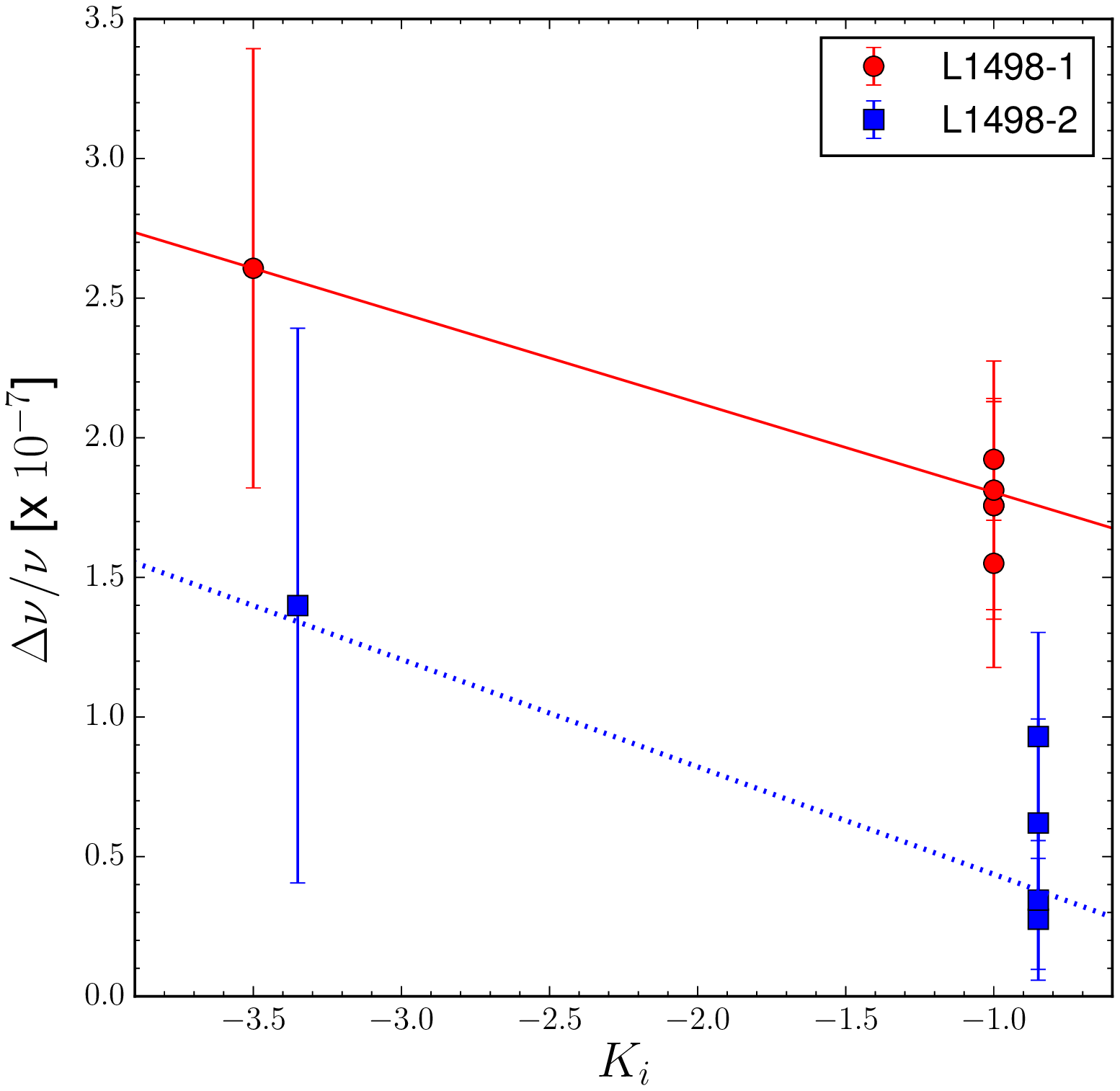}
    \caption{The relative shifts between the observed, after applying the velocity correction (see \mbox{Section \ref{sec:restframe_6}}), and the calculated frequencies from \protect \cite{Xu2008} of the methanol transitions observed in L1498-1 (red circles) and in L1498-2 (blue squares) are plotted against their sensitivity coefficients. The solid (red) and the dotted (blue) lines show the linear fit to the data set of L1498-1 and L1498-2, respectively, and their slopes represent the two values of $\dmm$ delivered. For clarity, a constant offset of $+0.15$ was applied to the measurements relative to L1498-2 on the x-axis.}
    \label{fig:mux_6}
\end{figure}

The velocity corrected frequencies of the detected methanol transitions were subsequently compared to the rest-frame frequencies via \mbox{Eq. (\ref{eq:shift_6})}. The comparison with the preferred rest frame frequencies as obtained from the \textsc{belgi} fit \citep{Xu2008} returned values of $\dmm_{1} = (-3.2 \pm 2.0_{\stat}) \times 10^{-8}$ and $\dmm_{2} = (-3.8 \pm 6.6_{\stat}) \times 10^{-8}$, for positions L1498-1 and L1498-2, respectively, as shown in \mbox{Fig. \ref{fig:mux_6}}. The larger uncertainty in the $\dmm$ value derived from L1498-2 reflects the spread in the observed frequencies of the lines with \mbox{$K_{i} = -1$}, which includes the shorter integration time spent on that position. Assuming that the physical conditions of the core are the same in L1498-1 and L1498-2 (see \mbox{Section \ref{sec:cloud_6}}), the weighted average of these two values yields to $\dmm = (-3.3 \pm 1.9_{\stat}) \times 10^{-8}$, hereafter the fiducial value of $\dmm$.

The comparison with the measured laboratory frequencies \citep{Tsunekawa1995,Muller2004} delivered the values of $\dmm_{1} = (3.6 \pm 14.5_{\stat}) \times 10^{-8}$ and $\dmm_{2} = (5.6 \pm 21.4_{\stat}) \times 10^{-8}$, for positions L1498-1 and L1498-2, respectively. The large statistical uncertainties on these values are due to the spread in the frequencies of the lines with $K_{i} = -1$, i.e. the 96 and the \mbox{145 GHz} lines, which reflects the inconsistencies between the experimental data sets.

\subsection{Hyperfine structure}
\label{subsubsec:hfs_6}
In the past two years the hyperfine structure in the microwave spectrum of methanol has attracted much attention \citep{Coudert2015,Belov2016,Lankhaar2016}, for a part connected to the importance of methanol as the most sensitive probe for varying constants. The underlying hyperfine structure of microwave transitions will cause an asymmetrical line shape, depending on the spread of the hyperfine components within the Doppler profile. However, the presently observed lines in L1498 do not display any asymmetry (see \mbox{Fig. \ref{fig:fits_6}}) and they can be very well reproduced by a single Gaussian shaped profile. 

In addition, unresolved hyperfine structure may cause a shift of the centre-of-gravity determined by the hyperfine-less rotational Hamiltonian. The hyperfine coupling involves three contributions: the magnetic dipole interactions between nuclei with spin $I > 0$, the coupling between the nuclear spins and the rotation, and the coupling between the nuclear spins and the torsion. While the spin-torsion interaction is represented by a vector, hence an irreducible tensor of \mbox{rank 1}, the contributions by the magnetic dipole interactions are described by \mbox{rank 2} tensors. The dipole-dipole interaction is represented by a traceless irreducible tensor of \mbox{rank 2} that retains the centre of gravity of the line, i.e. when weighting the intensity contributions of all hyperfine components. The spin-rotation coupling comprises also an irreducible tensor of \mbox{rank 2}, but in addition a scalar contribution (tensor of \mbox{rank 0}) giving rise to a trace that may shift the centre-of-gravity. This effect is quantified here based on the hyperfine calculations by \cite{Lankhaar2016}.

\begin{figure}
    \centering
    \includegraphics[width=\columnwidth]{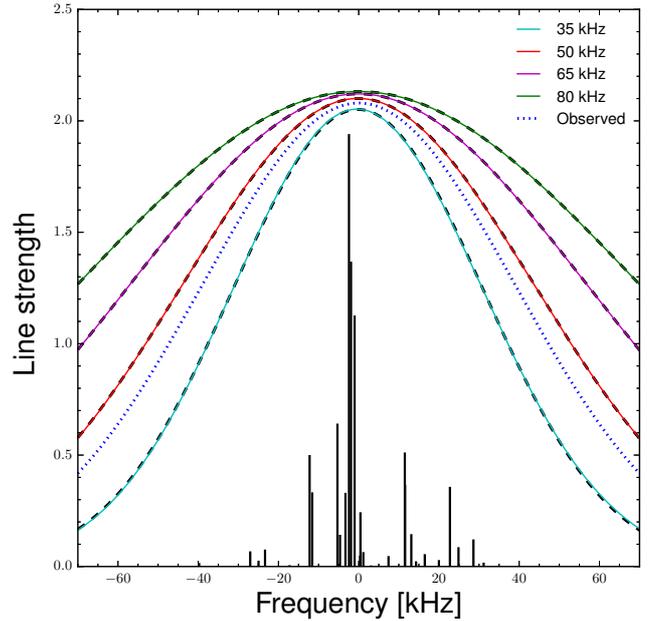}
    \caption{Comparison of the line shape derived from the hyperfine structure and calculated by \textsc{belgi} for the \mbox{$2_{0} \rightarrow 1_{0}$ A$^{+}$} transition at \mbox{$\nu_{\rest} = 96.741$ GHz}. The solid lines represent the convolution of the Gaussian profiles relative to the hyperfine components, while the dashed line represents the fit to the observed line profile centred on the calculated transition frequency. The dotted (blue) line corresponds to the fit to the observed line profile in L1498-1. The vertical bars indicate the individual hyperfine transitions with a length proportional to their intensity in the optically thin case under conditions of local thermodynamical equilibrium.}
    \label{fig:hyperfine_6}
\end{figure}

The calculated intensities for the individual hyperfine components of a transition \citep{Lankhaar2016} were convoluted using Gaussian profiles with \mbox{\emph{FWHM}$ = 35$-80 kHz}, as shown in \mbox{Fig. \ref{fig:hyperfine_6}} for the \mbox{$2_{0} \rightarrow 1_{0}$ A$^{+}$} transition, thus creating a line shape profile for each transition. This profile was subsequently compared and fitted to a Gaussian function and its computed line centre was compared with the position of the hyperfine-less zero-point. This procedure was followed for all transitions listed in \mbox{Table \ref{tab:rest_freq_6}}, resulting in small (\mbox{$< 1$ kHz}) frequency shifts of the rotational lines. It is noted that this shift comes in addition to the rotational structure and is not contained in the Hamiltonian underlying the \textsc{belgi} model, although some of the effects of hyperfine shift may, in principle, be covered by adapting the molecular parameters in the fitting procedure. 

However, since the shift introduced by the underlying hyperfine structure is below the kHz level, it is not expected to affect significantly the fiducial $\dmm$ values presented in this analysis. To estimate its impact on a varying $\mu$ and to account for possible moderate but not quantifiable deviations from the local thermodynamical equilibrium, an artificial shift of \mbox{$\pm 1$ kHz}, which is a conservative estimate, was introduced in the rest-frame frequency sets and new values for $\dmm$ were calculated using \mbox{Eq. (\ref{eq:shift_6})}. The difference with the  fiducial values was \mbox{$\delta^{\dmm}_{\textrm{\scriptsize{HFS}}} = 0.1 \times 10^{-8}$} at most, and this was added to the systematic error budget.  

\subsection{Doppler tracking}
\label{subsubsec:doppler_6}
The motion of the Earth around the Sun and Earth's rotation axis introduces Doppler shifts in the observed frequencies. The IRAM~30m telescope automatically corrects the observed frequencies, i.e. the band centre, for the velocity of the observatory with respect to the solar system barycentre. It is noted that correcting the band centre, introduces small frequency shifts to the positions of lines observed simultaneously at different frequencies, which are automatically corrected for in \textsc{class} (see \mbox{Section \ref{sec:obs_6}}). 

The sky frequencies were adjusted at the beginning of each scan, with a Doppler correction defined as:
\begin{equation}
    D = \frac{1}{1+\frac{\vtel}{c}},
    \label{eq:doppler_6}
\end{equation}
where $\vtel$ is the Doppler velocity of the telescope relative to the LSR, which is calculated using \mbox{Eq. (\ref{eq:optvel_6})}. As already outlined in \mbox{Section \ref{sec:obs_6}}, each scan consisted of an integration time of \mbox{3 min} off-source followed by \mbox{3 min} of integration on L1498. Thus, the methanol lines were measured starting \mbox{$\sim 3$ min} after the Doppler adjustment. It is noted that, while relative velocity shifts among the spectral lines introduce an effect that mimics a non-zero $\dmm$, a constant velocity shift does not have any impact on the fiducial value of $\dmm$, although it affects the $\vlsr$ value.

L1498 was observed on two consecutive days in July 2014, when its latitude in the ecliptic system was \mbox{$\sim 4$ deg}. This implies that the Earth was approaching the cloud with a very small acceleration, with respect to its orbit and hence with its almost maximal angular speed of \mbox{$\sim 10^{-5}$ \degs}. The motion of the Earth around the Sun causes, in a time interval of \mbox{$\Delta t = 3$ min}, a change in the velocity of the telescope relative to the LSR of \mbox{$\Delta \vtel \sim 1$ \ms} corresponding to a frequency shift of, at most, \mbox{0.6 kHz}. The variation of $\vtel$ during the off-source integration was not considered in the systematic error budget, as it affects all the methanol lines in the same way, while the $\Delta \vtel$ introduced during the integration on L1498 causes differential shifts of the methanol lines. Since the observations were performed at the same time in two consecutive days, these shifts were assumed to be in common for all scans. To quantify the impact of the Doppler tracking on the fiducial values, the observed frequencies were shifted by \mbox{$\pm 0.6$ kHz} and new $\dmm$ values were derived and compared with the fiducial ones. The largest offset from the fiducial values returned by this comparison is \mbox{$\delta^{\dmm}_{\textrm{\scriptsize{orbit}}} \le 0.1 \times 10^{-8}$}, which was added to the systematic error budget.

The Earth's rotation, the velocity of which is \mbox{$\sim 4.2 \times 10^{-3}$ \degs}, was estimated in a similar way. Throughout the observing run, the elevation of L1498 changed from a minimal value of \mbox{$30$ deg} to a maximal value of \mbox{$78$ deg}. The changes in the target's elevation result in velocity shifts in the range from $1.3$ to \mbox{$2.5$ \ms} for scans taken at the minimal and the maximal elevation value, respectively, during an integration time of \mbox{$3$ min}. Since both the off- and the on-source integrations deliver different velocity shifts, their effects were included in the systematic error budget. The total integration time of \mbox{6 min} delivered a spread of \mbox{$\sim 2.4$ \ms} in the telescope velocities due to the rotation of the Earth, which translates in a shift of, at most, \mbox{$\sim 1.2$ kHz} in the observed frequencies, which corresponds to a contribution to the systematic error budget of \mbox{$\delta^{\dmm}_{\textrm{\scriptsize{rotation}}} = 0.1 \times 10^{-8}$}. 

\subsection{Total systematic uncertainty}
\label{subsubsec:total_syst_6}
The total systematic uncertainty on the fiducial values was derived by adding in quadrature the contributions from the Doppler tracking and the underlying methanol hyperfine structure. This returned an error of \mbox{$\sim 0.2 \times 10^{-8}$}. Therefore, the final fiducial value of $\dmm$ obtained from the comparison of the observed frequencies with rest frame frequencies by \cite{Xu2008} is \mbox{$\dmm = (-3.3 \pm 1.9_{\stat} \pm 0.2_{\syst}) \times 10^{-8}$}.

\section{Conclusion}
\label{sec:conclusion_6}
Methanol (CH$_{3}$OH) emission in the cold core L1498 was investigated in order to constrain the dependence of the proton-to-electron mass ratio on matter density, thereby testing the chameleon scenario for theories beyond the Standard Model of physics \citep{Khoury2004,Brax2004}. The result presented in this study is derived through the methanol method, which involves observations of methanol transitions only, thereby avoiding assumptions of cospatiality among different molecules. This work strengthens the results previously obtained from radio astronomical searches \citep{Levshakov2010a,Levshakov2010b,Levshakov2013} by adding robustness against systematics effects such as chemical segregation. The methanol method provides future prospects for improved constraints on varying constants in the chameleon framework, if additional low-frequency lines can be detected.

In addition to this fundamental physics quest, the astrophysical properties of the L1498 cloud and its methanol A and E column densities were studied with Large Velocity Gradient radiative transfer calculations. The two spatially resolved methanol emission peaks (SE and NW) reported by \cite{Tafalla2006} were targeted for \mbox{$\sim 18$ hrs} using the IRAM 30m telescope. Six methanol transitions were detected in the brighter methanol emission peak, L1498-1, and five in the other position, L1498-2. 

All the lines were modelled using a single Gaussian profile and their observed frequencies were compared to rest-frame frequencies in order to derive a value for $\dmm$. Two sets of rest frame frequencies were considered: one including some high precision measurements \citep{Tsunekawa1995,Muller2004}, and one including calculated frequencies using the \textsc{belgi} code \citep{Xu1997,Xu2008}. These two sets are in good agreement, except for the frequencies of the transitions at \mbox{$\sim 145$ GHz} \citep{Tsunekawa1995}, which show offsets $> 30$ kHz between the measured and the calculated frequencies. In view of such an offset, the \textsc{belgi} outputs were preferred as the reference rest frame frequencies.

By comparing the accurate observed frequencies with the reference rest frame values reported by \cite{Xu2008}, a value of $\dmm$ was derived for each methanol emission peak. The weighted average of these values returned $\dmm = (-3.3 \pm 1.9_{\stat} \pm 0.2_{\syst}) \times 10^{-8}$, which is consistent with no variation at a level of $6 \times 10^{-8}$ ($3 \sigma$). This result is delivered by the analysis of emission lines belonging to a single molecular species, hence it is not depending on assumptions on the spatial distribution of the considered molecules. While \emph{E} and \emph{A} type methanol may be regarded as two different species, previous methanol observations in this system \citep{Tafalla2006} did not find evidence of chemical segregation between the two methanol types and our radiative transfer calculations indicate that they trace gas with similar physical properties. The $\dmm$ constraint is three times less stringent than that found by \cite{Levshakov2013}, which represents the most stringent constraint at the present day, but our limit is based on data from a single molecule and is therefore not affected by potential caveats caused by astrochemical processes.

The RADEX non-LTE model was used to derive the physical characteristics of the cloud. RADEX was used to find the best  solution by optimising the E- and A-type methanol column densities as functions of the kinetic temperature and the molecular hydrogen density. The best fit solutions for both L1498-1 and L1498-2 return molecular hydrogen densities of order \mbox{$n(\textrm{H}_{2}) \sim 3 \times 10^{5}$ cm$^{-3}$}, kinetic temperatures of \mbox{$\tkin \sim 6$ K} and total methanol column densities of several \mbox{$10^{12}$ cm$^{-2}$}. E- and A-type methanol abundance ratios appear to be close to unity.

The $\dmm$ value presented in this work can be improved by further observing methanol emission in cold cores. The $\dmm$ value presented here relies predominantly on the \mbox{$1_{0} - 1_{-1}$ E} transition at $\sim 157.270$ GHz, which is the only one with a sensitivity coefficient $K_{i} \ne -1$ in the considered sample. Observing extra lines with different sensitivities in L1498 will improve the $\Delta K_{i}$ and thereby the accuracy on $\dmm$. For example, observing the \mbox{$2_{0} \rightarrow 3_{-1}$ E} transition at \mbox{$\nu_{\rest} = 12.178$ GHz} (sensitivity coefficient of \mbox{$K_{i} = -32.5$}) will enhance the $\Delta K_{i}$ used to derive the $\dmm$ constraint, thereby improving it by one order of magnitude. This line has actually been observed in enhanced absorption again the cosmic microwave background radiation in the cold dark clouds TMC~1 and L~183 \citep{Walmsley1988}. Further observations of low frequency methanol transitions with larger sensitivity coefficients are planned with the Effelsberg 100m telescope and will improve the current constraint. Future observations of methanol lines in different cores showing low degrees of turbulence \citep[e.g. L1517B, ][]{Tafalla2004,Tafalla2006}, will provide independent constraints on the dependence of $\mu$ from matter density.

\section*{Acknowledgments}
The authors MD, HLB, and WU thank the Netherlands Foundation for Fundamental Research of Matter (FOM) for financial support. WU acknowledges the European Research Council for an ERC-Advanced grant under the European Union's Horizon 2020 research and innovation programme (grant agreement No 670168). AL is supported by the Russian Science Foundation (project 17-12-01256). This work was partially carried out within the Collaborative Research Council 956, subproject A6, funded by the Deutsche Forschungsgemeinschaft (DFG). We thank L.~H.~Xu (University of New Brunswick) for useful discussions about the calculated methanol rest frame frequencies. We thank A. v.d Avoird (Radboud University Nijmegen) and B. Lankhaar (Chalmers University G\"{o}teborg) for elucidating discussions and for making available the data on calculations of splittings and relative intensities of the hyperfine components for the six lines listed in Table \ref{tab:rest_freq_6}. We thank the staff of the IRAM 30m telescope for support during the observations.

\bibliographystyle{mnras}
\bibliography{bibliography}

\bsp
\label{lastpage}
\end{document}